\documentclass[12pt,british]{article}
\usepackage{a4}
\usepackage{amsmath}
\usepackage{a4}
\usepackage{amsmath}
\usepackage{babel}
\usepackage{enumerate}
\usepackage{epsfig}
\usepackage{float}
\usepackage{graphicx}

\begin{document}

\title{Nonequilibrium translational effects in evaporation and condensation}
\author{D.Bedeaux$^{1,2}$, S. Kjelstrup$^{1}$ and J.M. Rubi$^{1,3}$ \\
$^{1}$Department of Chemistry, Norwegian University of Science and\\
Technology, Trondheim, 7491 Trondheim, Norway \\
$^{2}$Leiden Institute of Chemistry, Leiden University,\\
PO Box 9502, 2300 RA Leiden, The Netherlands\\
$^{3}$Departament de Fisica Fonamental, Facultat de Fisica,\\
Universitat de Barcelona, Diagonal 647, 08028 Barcelona, Spain}
\maketitle
\date{}

\begin{abstract}
  This paper shows how mesoscopic nonequilibrium thermodynamics can be
  applied to condensation and evaporation. By extending the normal set
  of thermodynamic variables with two internal variables, we are able
  to give a new theoretical foundation for a mechanism of condensation
  that has been proposed from molecular simulation results. The flux
  does not follow a simple Arrhenius formula for small activation
  energies which are relevant here.  
\end{abstract}

\section{Introduction} 

Evaporation and condensation has been well studied, and many theories exist
for this important phenomenon. The theories are mostly rate theories.
Kinetic theory has long played a central role \cite{Ytrehus1997}. 
With information about the condensation
coefficient in this theory, the mass and heat fluxes are determined.
Recently Nagayama and Tsuruta \cite{Nagayama2003} have given a derivation of
the condensation coefficient using transition state theory
\cite{Eyring1965}. They based their theory on molecular dynamics
simulations of condensing
particles \cite{Tsuruta1999}, giving that the probability of a particle to
condense increases with an increasing normal component of the velocity of
the particle to the interface.

Common to rate theories \cite{Eyring1965} is that the driving forces of
transport do not appear explicitly in the expression for the rate.
Nonequilibrium thermodynamics gives, on the other hand, two independent
forces for the case of condensation or evaporation of a one-component
system. These are the chemical potential difference at constant temperature,
and the difference in inverse temperature \cite{Bedeaux1990, Rosjorde2001}.
In the nonequilibrium thermodynamic description of phase transition, the
heat and mass fluxes are linear in these forces.

Rubi and coworkers \cite{Vilar2001, Reguera1998} developed mesoscopic
nonequilibrium thermodynamics, following and generalizing the work described
in de Groot and Mazur \cite{deGroot1984}. This method, which describes the
system on the mesoscopic level, can give a more detailed thermodynamic
foundation for rate processes. This is done by including more variables than
are normal in a nonequilibrium thermodynamic description, for instance the
velocity of the particles, an important variable in rate theories. By
integrating the flux equations, which are now linear in the forces on the
mesoscopic level, nonlinear flux-force relations are obtained.

Rate theories and nonequilibrium thermodynamics aim to describe the same
phenomenon. It is therefore important to know the strength and weaknesses of
the theories, and under which conditions they become equivalent. It is the
main aim of this article to develop and apply mesocopic nonequilibrium
thermodynamics to evaporation and condensation. By developing the theory for
this case, we hope to strengthen the thermodynamic basis for this important
phenomenon. We shall present a detailed mesoscopic formulation, starting
with the macroscopic one, in order to define proper boundary conditions. As
new variables we introduce the condensation coordinate, called the reaction
coordinate by Nagayama and Tsuruta, and the normal velocity of the particles
as internal variables. By doing this we shall see that the earlier
description using nonequilibrium thermodynamics for interfaces \cite
{Bedeaux1990, Rosjorde2001} can be related to a richer, more detailed
formulation. The new level of description is also easier to compare to
kinetic theory. This theory will appear then as a special case.

The paper is organized as follows. The nonequilibrium theory for the
mesoscopic level is developed in section 2. We shall use the terminology
used in de Groot and Mazur \cite{deGroot1984}. Boundary conditions are given
for the fluxes along the internal coordinates. The result of the section is
a new expression for the excess entropy production rate. In the 3rd section
the description is restricted to stationary evaporation and condensation.
This simplifies the expression for the excess entropy production rate
considerably. The resulting flux equations are next given in section 4. For
large enough times one may neglect inertial effects due to acceleration of
particles. This leads to a simplified expression for the flux along the
condensation coordinate. Leaving the effects of the thermal force to future
work, we then proceed to neglect temperature differences across the
interface. Integrating the resulting expression for the flux along the
condensation coordinate over the internal variables, an expression is found
for the condensation flux across the interface in the isothermal case. Our
result for the condensation flux is discussed and compared to the
corresponding ones found using simple kinetic theory and in the work by
Nagayama and Tsuruta \cite{Nagayama2003}, in section 5. They are found to
reproduce the results of both these cases. In the final section some
conclusions are drawn.

\section{The system on the macroscopic level}

The system consists of a vapour in the presence of its liquid in a container
of volume $V$. The interface is planar. On a macroscopic scale (say in $\mu$%
m), the vapor is located in the region with $x<0$, while the liquid is
located in the region with $x>0$, see Fig.1a. The interface appears as a
discontinuity on this scale, but with excesses in entropy, internal energy,
and mass. This formulation is due to Gibbs \cite{Gibbs1961}. In the absence
of global equilibrium, there are gradients in temperature, $T$, and density, 
$\rho$, in the homogeneous liquid and vapour phases. Across the interface,
there is also a local variation in the same quantities, as well as in the
velocity distributions of the particles. This variation is, as Nagayama and
Tsuruta \cite{Nagayama2003} have shown, important for condensation.

The expression for the excess entropy production rate of the interface, in
the presence of a temperature difference and a difference in chemical
potential was given by Bedeaux and Kjelstrup \cite{Bedeaux1999b}. For a
one-component system we have: 
\begin{align}
\sigma ^{s}& =J_{q}^{\prime ,g}\left[ \frac{1}{T^{s}}-\frac{1}{T^{g}}\right]
+J_{q}^{\prime ,l}\left[ \frac{1}{T^{l}}-\frac{1}{T^{s}}\right]   \notag \\
& -J^{g}\left[ \frac{\mu ^{s}-\mu ^{g}\left( T^{s}\right) }{T^{s}}\right]
-J^{l}\left[ \frac{\mu ^{l}\left( T^{s}\right) -\mu ^{s}}{T^{s}}\right] 
\label{1}
\end{align}
The flux of heat from the vapour into the interface is $J_{q}^{\prime ,g}$ ,
while the one from the interface into the liquid is $J_{q}^{\prime ,l}$. The
corresponding mass fluxes are $J^{g}$ and $J^{l}$. When the independent
fluxes are the measurable heat fluxes and the mass fluxes on the vapor and
the liquid sides, the conjugate forces are the differences in inverse
temperatures (from the interface to the vapour, and from the liquid to the
interface) and the same differences for the chemical potentials at the
temperature of the interface divided by the temperature of the interface.
The superscripts $g,s$\ and $l$\ indicate values in the vapor, at the
interface and in the liquid, respectively. The chemical potential in the
interface is $\mu ^{s}$. When we write $\mu ^{g}\left( T^{s}\right) $ it
means that the chemical potential of the gas is taken at the temperature of
the interface. When the chemical potential of the gas is taken at the
temperature of the gas this is not separately indicated, thus $\mu ^{g}=\mu
^{g}\left( T^{g}\right) $. The frame of reference for the fluxes is the
interface. For simplicity we do not consider fluxes along the interface in
this paper.

\section{Mesoscopic nonequilibrium thermodynamics}

The purpose of this work is to provide a mesoscopic description of the
process, in terms of internal variables characterising the transition states
of the system, able to derive the kinetic law for the rate of condensation.

We shall replace the third and the fourth terms in Eq.(\ref{1}) by a more
detailed expression derived using two internal variables. The choice for
internal variables is motivated by the work of Nagayama and Tsuruta 
\cite{Nagayama2003}. 
We know from their work that the potential energy of condensing particles
has a maximum in the interfacial region and that the velocity component
normal to the interface is important for the likelihood of
condensation.  It is also known that the energy flux across the
interface has maximum  contribution from potential energy 
in the interface \cite{Ratkje1996}. In
order to describe this, we introduce a coordinate $\gamma $ for the degree
of condensation. This coordinate is similar to the degree of reaction in
transition state theory for chemical reactions \cite{Eyring1965}. By
introducing the degree of condensation as a variable, one enlarges the space
of the interface into an internal space, ranging from $\gamma =0$ to $\gamma
=1$. This is illustrated in Fig.1b. In other words, we expand the
interface in this coordinate (at $x=0)$. In the process, we must make sure that
the boundary conditions are correct.  

\begin{figure}
\begin{center}
\mbox{\psfig{file=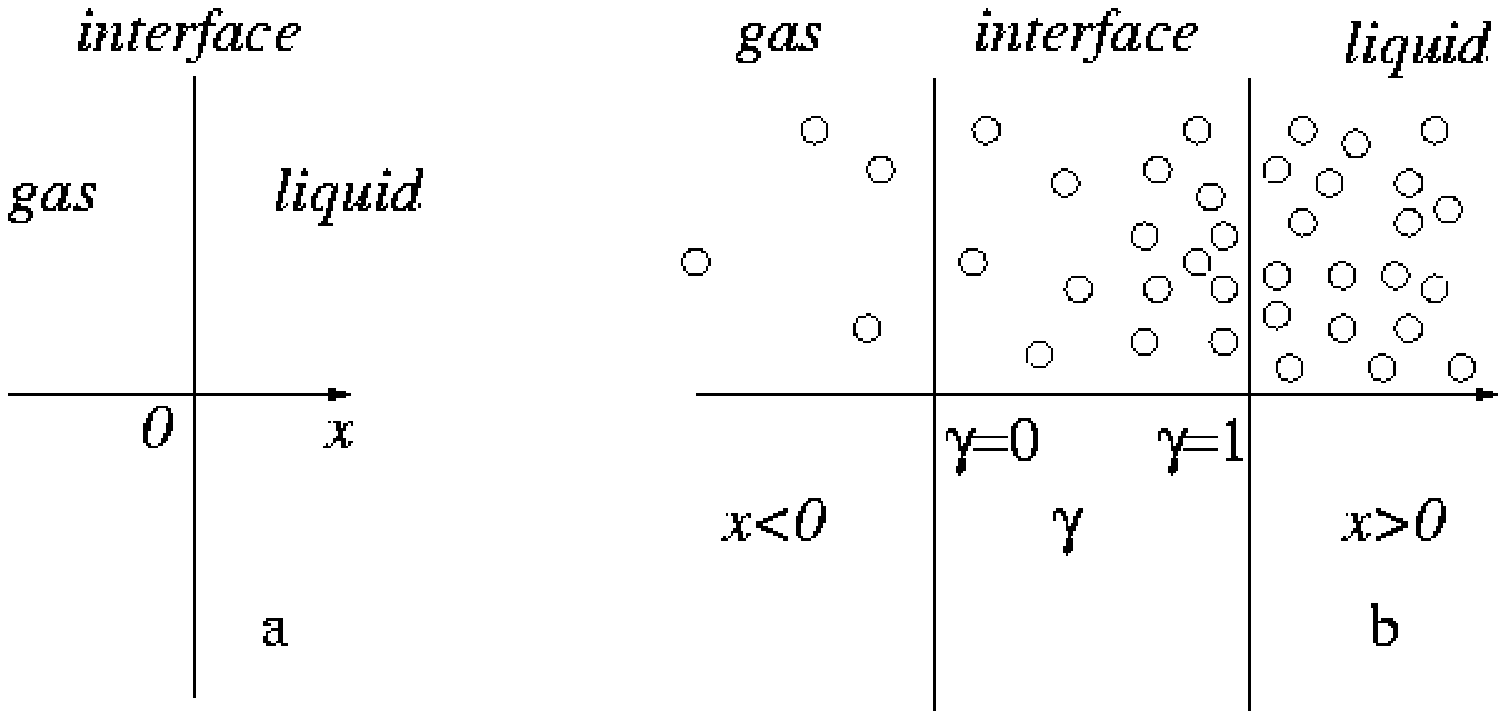,width=12cm}}
\caption{\label{fig:1}}
\end{center}
\end{figure}

In Fig.\ref{fig:1}a the chemical potential has a separate value in the
interface. The 
problem is to find the chemical potential in the interface as a
continuous function of the internal coordinate, i.e. to find the function $
\mu ^{s}(\gamma )$. We need boundary conditions for the chemical potential
that are compatible with Eq.(\ref{1}). The boundaries for the chemical
potential $\mu ^{s}(\gamma )$ in the $\gamma $-space are naturally: 
\begin{equation}
\frac{\mu ^{l}}{T^{l}}=\frac{\mu ^{s}(1)}{T^{s}}\text{ \ \ and \ \ }\frac{%
\mu ^{s}(0)}{T^{s}}=\frac{\mu ^{g}}{T^{g}}  \label{24a}
\end{equation}
Additionally, the flux along the condensation coordinate, $J_{\gamma
}^{s}(\gamma )$, must obey 
\begin{equation}
J_{\gamma }^{s}(0)=J^{g}\text{\ \ and\ \ \ }J_{\gamma }^{s}(1)=J^{l}
\label{15}
\end{equation}

We shall also use as internal coordinate, the component of the velocity
normal to the interface, $\mathrm{v}$ which varies from $
\mathrm{v}=-\infty$ to $\mathrm{v}=\infty$. The chemical potential will then
be a function of both $\gamma$ and $\mathrm{v}$. The relation between $%
\mu^{s}(\gamma)$ and $\mu^{s}(\gamma, \mathrm{v})$ will be given in the next
section. The flux along the internal coordinate ${\mathrm{v}}$, $J_{\mathrm{v%
}}^{s}$, is more commonly known as the acceleration of the particles. It
satisfies the following boundary condition: 
\begin{equation}
J_{\mathrm{v}}^{s}(\gamma ,-\infty)=0\text{\ \ and\ \ \ } J_{\mathrm{v}%
}^{s}(\gamma,\infty)=0
\end{equation}
We furthermore have 
\begin{equation}
J_{\gamma}^{s}(\gamma)\equiv \int_{-\infty}^{\infty}d\mathrm{v}\text{ }%
J_{\gamma}^{s}(\gamma,\mathrm{v})  \label{fl}
\end{equation}
This means that the value of $J_{\gamma}^{s}$\ integrated over all
velocities of the condensation coordinate is equal to the mass flux from the
vapor into the interface. There is no contribution to the mass flux at
infinitely high or low velocities at any coordinate $\gamma$.

\subsection{The chemical potential in internal variable space}

The variation in the excess entropy of the interface, $s^{s}$, in this
mesoscopic description is given by 
\begin{equation}
\delta s^{s}=\frac{1}{T^{s}}\delta u^{s}-\frac{1}{T^{s}}\int_{0}^{1}d\gamma%
\int_{-\infty}^{\infty}d\mathrm{v}\text{ }\mu^{s}(\gamma,\mathrm{v})\delta
f^{s}(\gamma,\mathrm{v})   \label{2}
\end{equation}
The excess entropy of the interface is a function of the excess internal
energy of the interface, $u^{s}$, and the probability, $f^{s}(\gamma,\mathrm{%
v}),$ to find a mole of particles in the interfacial area with the internal
variables $\gamma$\ and v, see \cite{Reguera1998}. As the equation
shows, the derivatives of the entropy with respect to these variables are
the inverse interface temperature, $1/T^{s}$, and the chemical potential, $%
\mu^{s} (\gamma,\mathrm{v})$, divided by the interface temperature, $T^{s}$.
In the general case, with $f^s(\gamma,\mathrm{v})$, there is a probability
for the particle to have a particular velocity. The velocity dependence is
integrated out: 
\begin{equation}
f^{s}(\gamma)\equiv\int_{-\infty}^{\infty}d\mathrm{v}\text{ }f^{s}(\gamma,%
\mathrm{v})   \label{5c}
\end{equation}
to give the probability distribution $f^s(\gamma)$. This distribution can be
interpreted as the probability to find a mole of particles in the
interfacial layer in Fig.1b with the internal variable $\gamma$ (in mol/m$^3$%
).

According to Gibbs entropy postulate \cite{deGroot1984}, we can write
that 
\begin{equation}
\mu ^{s}(\gamma ,\mathrm{v})=RT^{s}\ln f^{s}(\gamma ,\mathrm{v})+C^{s}\left(
\gamma ,\mathrm{v}\right)   \label{8}
\end{equation}
The potential function $C^{s}\left( \gamma ,\mathrm{v}\right) $, which
depends on $\gamma $\ and $\mathrm{v}$, may also depend on $T^{s}$. In
equilibrium the entropy has a maximum for a given value of the temperature
and it follows from Eq.(\ref{2}), that the chemical potential is constant.
We shall use the constant 
\begin{equation}
\mu _{eq}^{s}(\gamma ,\mathrm{v})=\mu ^{\ast }(T_{eq})-RT_{eq}\ln \left( 
\frac{2\pi RT_{eq}}{M}\right)   \label{8eq}
\end{equation}
The first term to the right, $\mu ^{\ast }(T_{eq})$, is the chemical
potential of the saturated vapor, considered as the standard state. This
chemical potential is independent of $\gamma $\ and $\mathrm{v}$. The last
term is minus two times the translational potential energy of one mole ideal
gas.

Alternatively, we can use the Gibbs postulate for the description where the
velocity coordinate has been integrated out. We then have 
\begin{equation}
\mu ^{s}(\gamma )=RT^{s}\ln f^{s}(\gamma )+C^{s}\left( \gamma \right) 
\label{8b}
\end{equation}
In equilibrium, we have 
\begin{equation}
\mu _{eq}^{s}(\gamma )=\mu ^{\ast }(T_{eq})-\frac{1}{2}RT_{eq}\ln \left( 
\frac{2\pi RT_{eq}}{M}\right)   \label{8beq}
\end{equation}
The second contribution is half the value of that in Eq.(\ref{8eq}). The
relation between the two expressions for the chemical potentials, Eqs.(\ref
{8eq}) and (\ref{8b}), is found after determining the values of the
functions $C^{s}\left( \gamma ,\mathrm{v}\right) $ and $C^{s}\left( \gamma
\right) $, see Eq.(\ref{5e}) below. Those values $C^{s}\left( \gamma ,%
\mathrm{v}\right) $ and $C^{s}\left( \gamma \right) $ will be obtained from
the probability distributions at equilibrium between liquid and vapour. 

\begin{figure}
\begin{center}
\mbox{\psfig{file=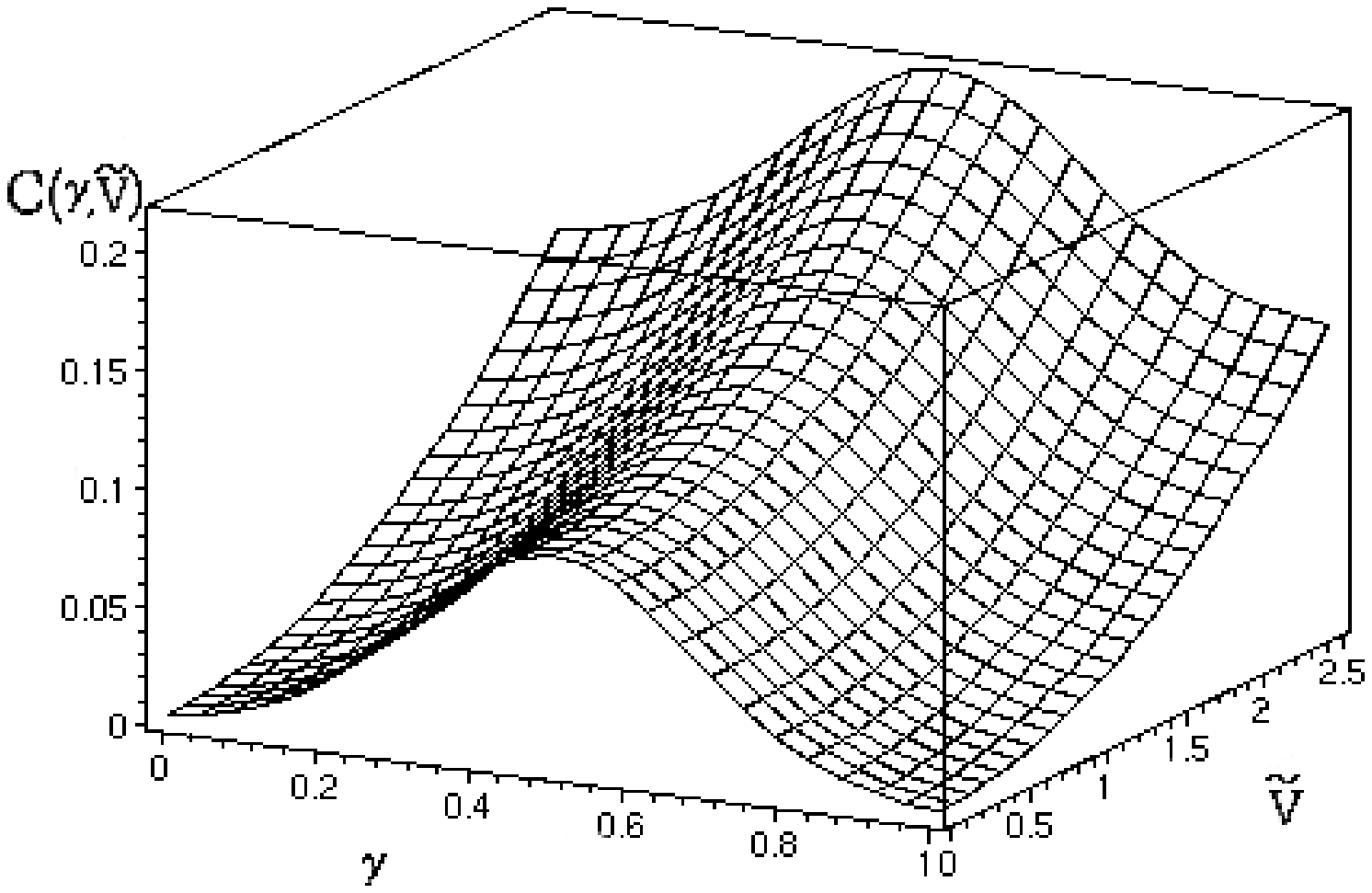,width=8cm}}
\caption{\label{fig:2}}
\end{center}
\end{figure}

Consider first the determination of $C^{s}\left( \gamma ,\mathrm{v}\right) $%
. The probability varies with the kinetic energy of the particle which for a
mole of particles, is given by $M\mathrm{v}^{2}/2$. We shall further assume
that the particles are crossing a potential energy barrier along the
condensation coordinate, $\phi (\gamma )$. The barrier is zero at the ends
of the condensation coordinate, $\phi (0)=\phi (1)=0$. In equilibrium, the
probability distribution is then given by: 
\begin{equation}
f_{eq}^{s}(\gamma ,\mathrm{v})=\exp \left[ \frac{\mu _{eq}^{s}(\gamma ,%
\mathrm{v})-\phi (\gamma )-\frac{1}{2}M\mathrm{v}^{2}}{RT_{eq}}\right] 
\label{9}
\end{equation}
By substituting this equation and Eq.(\ref{8eq}) into Eq.(\ref{8}), we find 
\begin{equation}
C^{s}\left( \gamma ,\mathrm{v}\right) =\phi (\gamma )+\frac{1}{2}M\mathrm{v}%
^{2}  \label{10}
\end{equation}
The function $C^{s}\left( \gamma ,\mathrm{v}\right)$ is illustrated in
Fig.\ref{fig:2}. This part of the chemical potential has a small maximum
along the $\gamma$-axis. Particles at the left of the barrier having higher velocities (and
consequently higher values of $C^{s}$) can easily cross the barrier in  $\gamma$-space going
through 
decreasing values of the velocity coordinate. This shows that the velocity distribution of the
particles plays an important role in condensation kinetics \cite{Nagayama2003}.


On the other hand, the probability distribution in the $\gamma $-space is: 
\begin{equation}
f_{eq}^{s}(\gamma )=\exp \left[ \frac{\mu _{eq}^{s}(\gamma )-\phi (\gamma )}{%
RT_{eq}}\right]   \label{9b}
\end{equation}
By substituting this equation and Eq.(\ref{8beq}) into Eq.(\ref{8b}), we
find 
\begin{equation}
C^{s}\left( \gamma \right) =\phi (\gamma )  \label{10b}
\end{equation}

The expressions are now in accordance with each other. This can be seen by
first substituting $f^{s}(\gamma,\mathrm{v})$\ and $f^{s}(\gamma)$, as
obtained from Eqs.(\ref{8}) and (\ref{8b}), into Eq.(\ref{5c}). The
following identity is obtained: 
\begin{equation}
\exp\left[ \frac{\mu^{s}(\gamma)-C^{s}\left( \gamma\right) }{RT^{s}}\right]
=\int_{-\infty}^{\infty}d\mathrm{v}\text{ }\exp\left[ \frac{\mu ^{s}(\gamma,%
\mathrm{v})-C^{s}\left( \gamma,\mathrm{v}\right) }{RT^{s}}\right] 
\label{5d}
\end{equation}
Dividing both sides by exp$\left[ -C^{s}\left( \gamma\right) /RT^{s}\right],$
and using Eqs.(\ref{10}) and (\ref{10b}), this can be written as 
\begin{equation}
\exp\left[ \frac{\mu^{s}(\gamma)}{RT^{s}}\right] =\int_{-\infty}^{\infty }d%
\mathrm{v}\text{ }\exp\left[ \frac{\mu^{s}(\gamma,\mathrm{v})-\frac{1} {2}M%
\mathrm{v}^{2}}{RT^{s}}\right]   \label{5e}
\end{equation}
Substitution of Eqs.(\ref{8eq}) and (\ref{8beq}) into this equation gives an
identity. This relation explains the factor 1/2 in front of the last term in
Eq.\ref{8beq}, while there is a 1 in the equation before it.

The boundary values for the chemical potential in $\gamma $-space, can now
be found using Eq.(\ref{9b}) and the expression for $\mu _{eq}^{s}$. We
obtain: 
\begin{align}
f_{eq}^{s}(0)& =\exp \left[ \frac{\mu _{eq}^{s}(0)}{RT_{eq}}\right]
=f_{eq}^{s}(1)=\exp \left[ \frac{\mu _{eq}^{s}(1)}{RT_{eq}}\right]   \notag
\\
& =\exp \left[ \frac{\mu ^{\ast }(T_{eq})}{RT_{eq}}-\frac{1}{2}\ln \left( 
\frac{2\pi RT_{eq}}{M}\right) \right] \equiv f^{\ast }(T_{eq})\sqrt{\frac{M}{%
2\pi RT_{eq}}}  \notag \\
p^{\ast }(T_{eq})& =RT_{eq}f^{\ast }(T_{eq})  \label{9cc}
\end{align}
The first equation says that the states with $\gamma =0$ and 1 are identical
at equilibrium, whereas the second equation assumes ideality.

In summary, we have given above two sets of chemical potential and
corresponding probability distribution. In the most general case, the
chemical potential is a continuous function of $\gamma ,\mathrm{v}$. In the
other case it is a function of $\gamma $. The relationship between $\mu
(\gamma ,\mathrm{v})$ and $\mu (\gamma )$ was given by integrating the most
general probability distribution function over $\mathrm{v}$-space. An
important consequence of the change in variables, is that the chemical
potential in the state of equilibrium changes. As the number of internal
variables increases the equilibrium state changes to lower values. With one
internal variable, the chemical potential is lowered by half the kinetic
energy of an ideal gas, with two internal variables it is lowered by twice
this value.

\subsection{The entropy production rate as a function of internal variables}

We can now follow the normal procedure and find the entropy production rate
in the variable space we are interested in. The procedure consists of
combining the time rate of change of Gibbs equation with the balance
equations. In the present case, there are three independent balance equations.

The balance equation for the probability distribution is 
\begin{equation}
\frac{\partial f^{s}(\gamma ,\mathrm{v})}{\partial t}=-\frac{\partial }{%
\partial \gamma }J_{\gamma }^{s}(\gamma ,\mathrm{v})-\frac{\partial }{%
\partial \mathrm{v}}J_{\mathrm{v}}^{s}(\gamma ,\mathrm{v})  \label{5}
\end{equation}
By integrating this equation over the velocity coordinate, we obtain the
balance equation for probability distribution along the condensation
coordinate alone 
\begin{equation}
\frac{\partial f^{s}(\gamma )}{\partial t}=-\frac{\partial }{\partial \gamma 
}J_{\gamma }^{s}(\gamma )  \label{5b}
\end{equation}
The conservation equation for internal energy is 
\begin{equation}
\frac{du^{s}}{dt}=J_{e}^{g}-J_{e}^{l}  \label{6}
\end{equation}
where each energy flux, $J_{e}^{g}$ or $J_{e}^{l}$, is a sum of the
measurable heat flux and the co-moving enthalpies, $J_{q}^{\prime }+hJ$, on
the vapor side and on the liquid side of the interface, respectively. The
balance for the entropy of the interface is 
\begin{equation}
\frac{ds^{s}}{dt}=J_{s}^{g}-J_{s}^{l}+\sigma ^{s}  \label{7}
\end{equation}
where the entropy flux is given by $J_{s}=(J_{e}-\mu J)/T=J_{q}^{\prime
}/T+sJ$\ in both bulk phases (with superscripts $g$\ and $l$, respectively). 

The time rate of change of the interfacial entropy density is obtained from
Eq.(\ref{2}): 
\begin{equation}
\frac{ds^{s}}{dt}=\frac{1}{T^{s}}\frac{du^{s}}{dt}-\frac{1}{T^{s}}%
\int_{0}^{1}d\gamma \int_{-\infty }^{\infty }d\mathrm{v}\text{ }\mu
^{s}(\gamma ,\mathrm{v})\frac{\partial f^{s}(\gamma ,\mathrm{v})}{\partial t}
\label{11}
\end{equation}
By substituting Eqs.(\ref{5}) and (\ref{6}) into this equation, we find 
\begin{align}
\frac{ds^{s}}{dt}& =\frac{1}{T^{s}}\left[ J_{e}^{g}-J_{e}^{l}\right]   \notag
\\
& +\frac{1}{T^{s}}\int_{0}^{1}d\gamma \int_{-\infty }^{\infty }d\mathrm{v}%
\text{ }\mu ^{s}(\gamma ,\mathrm{v})\left[ \frac{\partial }{\partial \gamma }%
J_{\gamma }^{s}(\gamma ,\mathrm{v})+\frac{\partial }{\partial \mathrm{v}}J_{%
\mathrm{v}}^{s}(\gamma ,\mathrm{v})\right]   \notag \\
& =\frac{1}{T^{s}}\left[ J_{q}^{\prime ,g}-J_{q}^{\prime ,l}\right] +\frac{1%
}{T^{s}}\left[ h^{g}J^{g}-h^{l}J^{l}\right]   \notag \\
& +\int_{-\infty }^{\infty }d\mathrm{v}\text{ }\frac{\mu ^{s}(1,\mathrm{v})}{%
T^{s}}J_{\gamma }^{s}(1,\mathrm{v})-\int_{-\infty }^{\infty }d\mathrm{v}%
\text{ }\frac{\mu ^{s}(0,\mathrm{v})}{T^{s}}J_{\gamma }^{s}(0,\mathrm{v}) 
\notag \\
& -\frac{1}{T^{s}}\int_{0}^{1}d\gamma \int_{-\infty }^{\infty }d\mathrm{v}%
\text{ }J_{\gamma }^{s}(\gamma ,\mathrm{v})\frac{\partial }{\partial \gamma }%
\mu ^{s}(\gamma ,\mathrm{v})  \notag \\
& -\frac{1}{T^{s}}\int_{0}^{1}d\gamma \int_{-\infty }^{\infty }d\mathrm{v}%
\text{ }J_{\mathrm{v}}^{s}(\gamma ,\mathrm{v})\frac{\partial }{\partial 
\mathrm{v}}\mu ^{s}(\gamma ,\mathrm{v})  \label{12}
\end{align}
where we have used partial integration and the fact that $J_{\mathrm{v}}^{s}$%
\ is zero for v = $\pm \infty $. Comparing this equation with Eq.(\ref{7})
we may identify the entropy production rate 
\begin{align}
\sigma ^{s}& =J_{q}^{\prime ,g}\left[ \frac{1}{T^{s}}-\frac{1}{T^{g}}\right]
+J_{q}^{\prime ,l}\left[ \frac{1}{T^{l}}-\frac{1}{T^{s}}\right]   \notag \\
& +\frac{1}{T^{s}}\left[ \mu ^{g}(T^{s})J^{g}-\mu ^{l}(T^{s})J^{l}\right]  
\notag \\
& +\int_{-\infty }^{\infty }d\mathrm{v}\text{ }\frac{\mu ^{s}(1,\mathrm{v})}{%
T^{s}}J_{\gamma }^{s}(1,\mathrm{v})-\int_{-\infty }^{\infty }d\mathrm{v}%
\text{ }\frac{\mu ^{s}(0,\mathrm{v})}{T^{s}}J_{\gamma }^{s}(0,\mathrm{v}) 
\notag \\
& -\frac{1}{T^{s}}\int_{0}^{1}d\gamma \int_{-\infty }^{\infty }d\mathrm{v}%
\text{ }J_{\gamma }^{s}(\gamma ,\mathrm{v})\frac{\partial }{\partial \gamma }%
\mu ^{s}(\gamma ,\mathrm{v})  \notag \\
& -\frac{1}{T^{s}}\int_{0}^{1}d\gamma \int_{-\infty }^{\infty }d\mathrm{v}%
\text{ }J_{\mathrm{v}}^{s}(\gamma ,\mathrm{v})\frac{\partial }{\partial 
\mathrm{v}}\mu ^{s}(\gamma ,\mathrm{v})  \label{13}
\end{align}
In order to simplify this equation we need boundary conditions for $\gamma =0
$ and $1$. For $\gamma =0$\ the particles leave the vapor and start moving
along the condensation coordinate. At $\gamma =1$\ the particles are at the
end of the condensation coordinate and enter the liquid. In the vapor and
the liquid we do not use the velocity coordinate as an internal variable.
This corresponds to the assumption that the chemical potential in those
regions does not depend on the velocity. At the ends of the condensation
coordinate, we shall similarly assume that the chemical potential is
independent of the velocity. In agreement with our previous choice for
standard states, we have using Eq.(\ref{5e}): 
\begin{equation}
\mu ^{s}(0,\mathrm{v})+\frac{1}{2}RT^{s}\ln \left( \frac{2\pi RT^{s}}{M}%
\right) =\mu ^{s}(0)\text{ \ \ and\ \ \ }\mu ^{s}(1,\mathrm{v})+\frac{1}{2}%
RT^{s}\ln \left( \frac{2\pi RT^{s}}{M}\right) =\mu ^{s}(1)  \label{14}
\end{equation}

Substitution of Eq.(\ref{14}) into Eq.(\ref{13}) results in 
\begin{align}
\sigma ^{s}& =J_{q}^{\prime ,g}\left[ \frac{1}{T^{s}}-\frac{1}{T^{g}}\right]
+J_{q}^{\prime ,l}\left[ \frac{1}{T^{l}}-\frac{1}{T^{s}}\right] +\frac{1}{2}%
R\left( J^{g}-J^{l}\right) \ln \left( \frac{2\pi RT^{s}}{M}\right)   \notag
\\
& -J^{g}\frac{\mu ^{s}(0)-\mu ^{g}(T^{s})}{T^{s}}-J^{l}\frac{\mu
^{l}(T^{s})-\mu ^{s}(1)}{T^{s}}  \notag \\
& -\frac{1}{T^{s}}\int_{0}^{1}d\gamma \int_{-\infty }^{\infty }d\mathrm{v}%
\text{ }J_{\gamma }^{s}(\gamma ,\mathrm{v})\frac{\partial }{\partial \gamma }%
\mu ^{s}(\gamma ,\mathrm{v})  \notag \\
& -\frac{1}{T^{s}}\int_{0}^{1}d\gamma \int_{-\infty }^{\infty }d\mathrm{v}%
\text{ }J_{\mathrm{v}}^{s}(\gamma ,\mathrm{v})\frac{\partial }{\partial 
\mathrm{v}}\mu ^{s}(\gamma ,\mathrm{v})  \label{16}
\end{align}
The first two terms are identical to the first two terms in Eq.(\ref{1}).
When $\mu ^{s}(\gamma ,\mathrm{v})=\mu ^{s}$ is constant, the last two terms
are zero and the fourth and fifth term are identical to the third and fourth
term in Eq.(\ref{1}). The third term is not independent of the two
following. It is zero in the stationary state, however. The last two terms
provide the new element in this mesoscopic description.

\section{Stationary evaporation and condensation}

Most experiments, Monte Carlo and molecular dynamics simulations are
performed for stationary evaporation and condensation. As the description of
this case simplifies considerably, we shall restrict ourselves to this case.
For the stationary case it follows from the balance equations that 
\begin{equation}
J_{e}^{g}=J_{e}^{l}\equiv J_{e}\text{ \ \ and \ \ \ }J^{g}=J^{l}\equiv J
\label{17}
\end{equation}
For the measurable heat fluxes, this relation gives 
\begin{equation}
J_{q}^{\prime ,l}=J_{q}^{\prime ,g}+J\left( h^{l}-h^{g}\right)   \label{18}
\end{equation}
Furthermore it follows from Eq.(\ref{5}), using stationarity, that 
\begin{equation}
\frac{\partial }{\partial \gamma }J_{\gamma }^{s}(\gamma ,\mathrm{v})+\frac{%
\partial }{\partial \mathrm{v}}J_{\mathrm{v}}^{s}(\gamma ,\mathrm{v})=0
\label{19}
\end{equation}
Integrating this equation over v, using $J_{\mathrm{v}}^{s}(\gamma ,-\infty
)=J_{\mathrm{v}}^{s}(\gamma ,\infty )=0$\ and Eq.(\ref{15}), gives 
\begin{equation}
\int_{-\infty }^{\infty }d\mathrm{v}\text{ }J_{\gamma }^{s}(\gamma ,\mathrm{v%
})=J_{\gamma }^{s}(\gamma )=J  \label{20}
\end{equation}

Using these relations the excess entropy production rate may be written as 
\begin{align}
\sigma ^{s}& =J_{q}^{\prime ,g}\left[ \frac{1}{T^{l}}-\frac{1}{T^{g}}\right]
+J\left( h^{l}-h^{g}\right) \left[ \frac{1}{T^{l}}-\frac{1}{T^{s}}\right]  
\notag \\
& -\frac{J}{T^{s}}\left[ \mu ^{l}(T^{s})-\mu ^{s}(1)+\mu ^{s}(0)-\mu
^{g}(T^{s})\right]   \notag \\
& -\frac{1}{T^{s}}\int_{0}^{1}d\gamma \int_{-\infty }^{\infty }d\mathrm{v}%
\text{ }J_{\gamma }^{s}(\gamma ,\mathrm{v})\frac{\partial }{\partial \gamma }%
\mu ^{s}(\gamma ,\mathrm{v})  \notag \\
& -\frac{1}{T^{s}}\int_{0}^{1}d\gamma \int_{-\infty }^{\infty }d\mathrm{v}%
\text{ }J_{\mathrm{v}}^{s}(\gamma ,\mathrm{v})\frac{\partial }{\partial 
\mathrm{v}}\mu ^{s}(\gamma ,\mathrm{v})  \notag \\
& =J_{q}^{\prime ,g}\left[ \frac{1}{T^{l}}-\frac{1}{T^{g}}\right] -J\left[ 
\frac{\mu ^{l}}{T^{l}}-\frac{\mu ^{s}(1)}{T^{s}}+\frac{\mu ^{s}(0)}{T^{s}}-%
\frac{\mu ^{g}(T^{l})}{T^{l}}\right]   \notag \\
& -\frac{1}{T^{s}}\int_{0}^{1}d\gamma \int_{-\infty }^{\infty }d\mathrm{v}%
\text{ }J_{\gamma }^{s}(\gamma ,\mathrm{v})\frac{\partial }{\partial \gamma }%
\mu ^{s}(\gamma ,\mathrm{v})  \notag \\
& -\frac{1}{T^{s}}\int_{0}^{1}d\gamma \int_{-\infty }^{\infty }d\mathrm{v}%
\text{ }J_{\mathrm{v}}^{s}(\gamma ,\mathrm{v})\frac{\partial }{\partial 
\mathrm{v}}\mu ^{s}(\gamma ,\mathrm{v})  \label{21}
\end{align}
where we used Eqs.(\ref{14}) and (\ref{20}). In the last identity we have
also employed the thermodynamic identity 
\begin{equation}
\frac{\partial (\mu /T)}{\partial (1/T)}=\mu -T\frac{\partial \mu }{\partial
T}=\mu +Ts=h  \label{22}
\end{equation}
where the differentials are evaluated at constant pressure. The pressure is
constant because the system is in mechanical equilibrium.

If one uses the energy flux rather than the heat flux in Eq.(\ref{21}) it
becomes 
\begin{align}
\sigma ^{s}& =J_{e}\left[ \frac{1}{T^{l}}-\frac{1}{T^{g}}\right] -J\left[ 
\frac{\mu ^{l}}{T^{l}}-\frac{\mu ^{s}(1)}{T^{s}}+\frac{\mu ^{s}(0)}{T^{s}}-%
\frac{\mu ^{g}}{T^{g}}\right]   \notag \\
& -\frac{1}{T^{s}}\int_{0}^{1}d\gamma \int_{-\infty }^{\infty }d\mathrm{v}%
\text{ }J_{\gamma }^{s}(\gamma ,\mathrm{v})\frac{\partial }{\partial \gamma }%
\mu ^{s}(\gamma ,\mathrm{v})  \notag \\
& -\frac{1}{T^{s}}\int_{0}^{1}d\gamma \int_{-\infty }^{\infty }d\mathrm{v}%
\text{ }J_{\mathrm{v}}^{s}(\gamma ,\mathrm{v})\frac{\partial }{\partial 
\mathrm{v}}\mu ^{s}(\gamma ,\mathrm{v})  \label{23}
\end{align}
In this form it is clear that in order to transfer a particle from the vapor
to the liquid, it must first be transferred from the vapor to the beginning
of the condensation coordinate, then it has to go along this coordinate to
its end and finally the particle should be transferred to the liquid. The
total resistance to go from one end to the other is the sum of the
resistances for these three steps. The description without internal
coordinates is obtained by setting $\mu ^{s}(\gamma ,\mathrm{v})=\mu ^{s}$\
in Eq.(\ref{23}). In that case only the first two terms remain. The force
giving mass transfer has simplified to $\mu ^{l}/T^{l}-\mu ^{g}/T^{g}$. We
see that the meaning of introduction of internal variables is to distribute
this force along the condensation coordinate. In line with the introduction
of the internal coordinates Eq.(\ref{24a}), is a natural choice to make. This
then simplifies Eq.(\ref{23}) to 
\begin{align}
\sigma ^{s}& =J_{e}\left[ \frac{1}{T^{l}}-\frac{1}{T^{g}}\right] -\frac{1}{%
T^{s}}\int_{0}^{1}d\gamma \int_{-\infty }^{\infty }d\mathrm{v}\text{ }%
J_{\gamma }^{s}(\gamma ,\mathrm{v})\frac{\partial }{\partial \gamma }\mu
^{s}(\gamma ,\mathrm{v})  \notag \\
& -\frac{1}{T^{s}}\int_{0}^{1}d\gamma \int_{-\infty }^{\infty }d\mathrm{v}%
\text{ }J_{\mathrm{v}}^{s}(\gamma ,\mathrm{v})\frac{\partial }{\partial 
\mathrm{v}}\mu ^{s}(\gamma ,\mathrm{v})  \label{25}
\end{align}

\section{Flux equations}

The stationary state excess entropy production rate that was derived above
has three fluxes and forces. Nonequilibrium thermodynamic theory prescribes
that each flux is a linear combination of each force of the system. Using
Eq.(\ref{23}), the linear relation for the fluxes in terms of the forces
become

\begin{align}
& J_{e}=L_{ee}\Delta(\frac{1}{T})-\int d\gamma d\mathrm{v}L_{e\gamma}(\gamma,%
\mathrm{v})\frac{\partial}{\partial\gamma}\frac{\mu^{s}(\gamma ,\mathrm{v})}{%
T^{s}}-\int d\gamma d\mathrm{v}L_{e\mathrm{v}}(\gamma ,\mathrm{v})\frac{%
\partial}{\partial\mathrm{v}}\frac{\mu^{s}(\gamma ,\mathrm{v})}{T^{s}} 
\notag \\
& J_{\gamma}(\gamma,\mathrm{v})=L_{\gamma e}(\gamma,\mathrm{v})\Delta (\frac{%
1}{T})-L_{\gamma\gamma}(\gamma,\mathrm{v})\frac{\partial}{\partial\gamma}%
\frac{\mu^{s}(\gamma,\mathrm{v})}{T^{s}}-L_{\gamma\mathrm{v}}(\gamma,\mathrm{%
v})\frac{\partial}{\partial\mathrm{v}}\frac{\mu^{s}(\gamma,\mathrm{v})}{T^{s}%
}  \notag \\
& J_{\mathrm{v}}(\gamma,\mathrm{v})=L_{\mathrm{v}e}(\gamma,\mathrm{v})\Delta(%
\frac{1}{T})-L_{\mathrm{v}\gamma}(\gamma,\mathrm{v})\frac{\partial }{%
\partial\gamma}\frac{\mu^{s}(\gamma,\mathrm{v})}{T^{s}}-L_{\mathrm{vv}%
}(\gamma,\mathrm{v})\frac{\partial}{\partial\mathrm{v}}\frac{\mu^{s}(\gamma,%
\mathrm{v})}{T^{s}}   \label{4.1}
\end{align}
where $\Delta(1/T)=1/T^{l}-1/T^{g}$. All conductivity coefficients $L_{ij}$
except $L_{ee}$\ are functions of the internal coordinates, while they all
depend on $T^{s}$\ and $c^{s}$. The matrix of conductivity coefficients
satisfies the Onsager symmetry relations.

\subsection{Mechanical equilibrium}

A non-zero flux of particles in velocity-space, means that there is an
acceleration, which is partially balanced by inertial effects. In the
absence of acceleration, $J_{\mathrm{v}}$=0, and the system is in mechanical
equilibrium. Equation (\ref{4.1}) gives: 
\begin{equation}
L_{\mathrm{v}e}(\gamma ,\mathrm{v})\Delta (\frac{1}{T})-L_{\mathrm{v}\gamma
}(\gamma ,\mathrm{v})\frac{\partial }{\partial \gamma }\frac{\mu ^{s}(\gamma
,\mathrm{v})}{T^{s}}-L_{\mathrm{vv}}(\gamma ,\mathrm{v})\frac{\partial }{%
\partial \mathrm{v}}\frac{\mu ^{s}(\gamma ,\mathrm{v})}{T^{s}}=0  \label{4.2}
\end{equation}
The last term, that contains a main coefficient, is a dissipative term (the
intertial term), while the two first terms represent work terms. The central
term is the work done by moving a mole of particles up the barrier in the
internal variable space. We can use this relation to eliminate $\partial \mu
^{s}(\gamma ,\mathrm{v})/\partial \mathrm{v}$\ from the other fluxes. This
gives for the flux along the condensation coordinate

\begin{equation}
J_{\gamma}(\gamma,\mathrm{v})= \mathsf{L}_{\gamma e}(\gamma,\mathrm{v}
)\Delta(\frac{1}{T})-\mathsf{L}_{\gamma\gamma}(\gamma,\mathrm{v} )\frac{%
\partial}{\partial\gamma}\frac{\mu^{s}(\gamma,\mathrm{v})}{T^{s} } 
\label{4.5}
\end{equation}
where 
\begin{equation}
\mathsf{L}_{\gamma e}(\gamma,\mathrm{v})=L_{\gamma e}(\gamma,\mathrm{v})-%
\frac{L_{\gamma\mathrm{v}}(\gamma,\mathrm{v})L_{\mathrm{v}e}(\gamma ,\mathrm{%
v})}{L_{\mathrm{vv}}(\gamma,\mathrm{v})}\text{ ,\ \ \ }\mathsf{L}%
_{\gamma\gamma}(\gamma,\mathrm{v})=L_{\gamma\gamma}(\gamma ,\mathrm{v})-%
\frac{L_{\gamma\mathrm{v}}(\gamma,\mathrm{v})L_{\mathrm{v}\gamma }(\gamma,%
\mathrm{v})}{L_{\mathrm{vv}}(\gamma,\mathrm{v})}   \label{4.6}
\end{equation}

\subsection{The isothermal case}

When the system in mechanical equilibrium is also isothermal, Eq.(\ref{4.5})
further reduces to 
\begin{equation}
J_{\gamma }(\gamma ,\mathrm{v})=-\mathsf{L}_{\gamma \gamma }(\gamma ,\mathrm{%
v})\frac{\partial \mu ^{s}(\gamma ,\mathrm{v})}{T\partial \gamma }
\label{4.8}
\end{equation}
where $T$\ is the uniform temperature. It is now convenient to write the
gradient of the chemical potential in an alternative form. Using Eqs.(\ref{8}%
) and (\ref{10}), we have 
\begin{equation}
\mu ^{s}(\gamma ,\mathrm{v})=RT^{s}\ln f^{s}(\gamma ,\mathrm{v})+\phi
(\gamma )+\frac{1}{2}M\mathrm{v}^{2}=RT\ln \left[ f^{s}(\gamma ,\mathrm{v}%
)f_{\gamma }^{s}(\gamma )f_{\text{v}}^{s}(\text{v})\right]   \label{4.3}
\end{equation}
where we defined 
\begin{equation}
RT\ln f_{\gamma }^{s}(\gamma )\equiv \phi (\gamma )\text{ \ \ and \ \ }RT\ln
f_{\text{v}}^{s}(\text{v})\equiv \frac{1}{2}M\mathrm{v}^{2}  \label{4.4}
\end{equation}
Using now Eq. (\ref{4.3}) we write Eq.(\ref{4.8}) as 
\begin{equation}
J_{\gamma }(\gamma ,\mathrm{v})=-\frac{\mathsf{L}_{\gamma \gamma }(\gamma ,%
\mathrm{v})R}{f^{s}(\gamma ,\mathrm{v})f_{\gamma }^{s}(\gamma )f_{\text{v}%
}^{s}(\text{v})}\frac{\partial }{\partial \gamma }\exp \frac{\mu ^{s}(\gamma
,\mathrm{v})}{RT}  \label{4.10}
\end{equation}
It is commonly found that the conductance is proportional to $f^{s}(\gamma ,%
\mathrm{v})$, so we will assume that the diffusion coefficient along the
condensation coordinate, which we define by 
\begin{equation}
D(\text{v})\equiv \frac{L_{\gamma \gamma }(\gamma ,\mathrm{v})R}{%
f^{s}(\gamma ,\mathrm{v})}  \label{4.12}
\end{equation}
is independent of $\gamma $. Equation(\ref{4.10}) can then be written in the
form 
\begin{equation}
f_{\gamma }^{s}(\gamma )J_{\gamma }(\gamma ,\mathrm{v})=-\frac{D(\text{v})}{%
f_{\text{v}}^{s}(\text{v})}\frac{\partial }{\partial \gamma }\exp \frac{\mu
^{s}(\gamma ,\mathrm{v})}{RT}  \label{4.13}
\end{equation}
Integration of this equation over $\gamma $ gives, using Eq.(\ref{24a}), 
\begin{equation}
\int_{0}^{1}d\gamma f_{\gamma }^{s}(\gamma )J_{\gamma }(\gamma ,\mathrm{v})=-%
\frac{D(\text{v})}{f_{\text{v}}^{s}(\text{v})}\left( \exp \frac{\mu ^{l}}{RT}%
-\exp \frac{\mu ^{g}}{RT}\right)   \label{4.14}
\end{equation}
Further integration of this equation over $\mathrm{v}$, taking
Eq.(\ref{20}) into account, yields: 
\begin{equation}
J=-\left[ \int_{0}^{1}d\gamma f_{\gamma }^{s}(\gamma )\right] ^{-1}\int d%
\text{v }D(\text{v})\exp \left[ -\frac{M\text{v}^{2}}{2RT}\right] \left(
\exp \frac{\mu ^{l}}{RT}-\exp \frac{\mu ^{g}}{RT}\right)   \label{4.11}
\end{equation}

For the chemical potential of the vapor we have, using Eqs.(\ref{24a}) and (
\ref{9cc}), 
\begin{equation}
\mu ^{g}=\mu ^{s}(0)=\mu ^{\ast }(T)-\frac{1}{2}RT\ln \left( \frac{2\pi RT}{M%
}\right) +RT\ln \frac{p}{p^{\ast }(T)}  \label{4.16}
\end{equation}
The chemical potential of the liquid corresponds to the chemical potential
of the saturated vapor at the temperature of the liquid:

\begin{equation}
\mu ^{l}=\mu ^{s}(1)=\mu ^{\ast }(T)-\frac{1}{2}RT\ln \left( \frac{2\pi RT}{M%
}\right)   \label{4.17}
\end{equation}
Substitution of these expressions in Eq.(\ref{4.11}), and using Eq.(\ref{9cc}%
) results in 
\begin{equation}
J=\left[ \int_{0}^{1}d\gamma e^{\phi /RT}\right] ^{-1}\int d\text{v }\frac{D(%
\text{v})}{RT}\exp \left[ -\frac{M\text{v}^{2}}{2RT}\right] \sqrt{\frac{M}{%
2\pi RT}}\text{ }\left[ p-p^{\ast }(T)\right]   \label{4.18}
\end{equation}
We define the first integral to the right as 
\begin{equation}
\exp \left( -\frac{E_{0}}{RT}\right) =\left[ \int_{0}^{1}d\gamma f_{\gamma }^{s}(\gamma )\right] ^{-1}
\end{equation}
where we $E_{0}$\ is an activation energy. The final result for isothermal
phase transition at mechanical equilibrium is then: 
\begin{equation}
J=\exp \left( -\frac{E_{0}}{RT}\right) \int d\text{v }\frac{D(\text{v})}{RT}%
\exp \left[ -\frac{M\text{v}^{2}}{2RT}\right] \sqrt{\frac{M}{2\pi RT}}\text{ 
}\left[ p-p^{\ast }(T)\right]   \label{4.18b}
\end{equation}

The flux of mass according to this equation is driven by the difference
between the vapour pressure and the saturation pressure. It is proportional
to an integral that depends exponentially upon the velocity of the particles
that are hitting the interface, as well as a diffusion coefficient in
velocity space. The flux depends also on an activation energy $E_0$, in a
common way.

\section{Discussion}

The theory we have presented can in general be applied to cases that are
away from mechanical equilibrium and are not isothermal. Under mechanical
and isothermal conditions it gives expressions for the condensation rate
which agree with those obtained from kinetic theory and transition state
theory. We will proceed in this section to show that agreement and will
leave for the next section a general discussion to summarise the main
characteristics and to point out some possible extensions.

\subsection{The simplest version of kinetic theory}

Kinetic theory provides expressions for the unidirectional rates of
evaporation or condensation, and the net rate as a difference between
these. It establishes that the evaporation rate is proportional to the
pressure of the saturated vapour (which gives the state of the
liquid), while the condensation rate is proportional to the actual
vapour pressure \cite{Ytrehus1997}. This gives as maximum
condensation rate for isothermal conditions:
\begin{equation}
J=\left( 2\pi MRT\right) ^{-1/2}\text{ }\left[ p-p^{\ast }(T)\right] 
\label{5.2}
\end{equation}

In the most simple kinetic theory, there is no potential barrier  along the
reaction coordinate ($E_0=0$). In our formulation this means that $f_{\gamma }^{s}(\gamma )=1$ in our
expressions. We then reproduce the result from kinetic theory by using: 
\begin{equation}
D(\text{v})=\text{v }  \label{5.1}
\end{equation}
Substituting this into Eq.(\ref{4.18}) and restricting the integration to
positive velocities one obtains Eq.(\ref{5.2}). 

If one takes a potential barrier into account, as we will do below, we
obtain a prefactor in Eq.(\ref{5.2}). The prefactor, which is the
condensation coefficient, gives the fraction of incident particles
that condenses. The equation then becomes equal to the Hertz-Knudsen
formula for isothermal conditions \cite{Hertz1882,Knudsen1915}. In
Schrage's formulation \cite{Schrage1953}, the
Maxwellian distribution of the incident particles is corrected by a
possible net bulk velocity. This type of correction will not be
considered now. Neither will we consider more recent results of kinetic
theory.

\subsection{Modified transition state theory}

Nagayama and Tsuruta \cite{Nagayama2003} found that the condensation
coefficient depended on the incident velocity of the particles. Their
system was in equilibrium and had no net bulk velocity. They
calculated the activation energy as the difference in the mean kinetic energy
of condensing and Maxwellian particles.

If there is a potential barrier along the condensation
coordinate, we have $f_{\gamma }^{s}(\gamma )\neq 1.$ This also affects
the expression for the diffusion coefficient $D($v$)$. The expression
for the condensation flux then becomes
\begin{equation}
J=\sigma _{c}\left( T\right) \left( 2\pi MRT\right) ^{-1/2}\text{ }\left[
p-p^{\ast }(T)\right]   \label{5.3}
\end{equation}
where the prefactor, the condensation coefficient, is given by 
\begin{equation}
\sigma _{c}\left( T\right) =  \exp
\left( -\frac{E_{0}}{RT} \right) \int_{0}^{\infty} d\text{v }\frac{MD(\text{v})}{RT}\exp \left[
-\frac{M\text{v}^{2}}{2RT}\right]   \label{5.4}
\end{equation}

The dependence of the fraction of condensing particles on the
incident velocity \cite{Nagayama2003}, modifies the expression
for $D($v$)$\ to
\begin{equation}
D(\text{v})=\text{v }\left( 1-\beta \exp \left[ -\frac{M\text{v}^{2}}{2RT}%
\right] \right)   \label{5.5}
\end{equation}
For the
condensation coefficient we then obtain
\begin{equation}
\sigma _{c}\left( T\right)= \exp
\left( -\frac{E_{0}}{RT}\right) \left( 1-\frac{\beta }{2}\right) 
\label{5.6}
\end{equation}
Following Nagayama and Tsuruta \cite{Nagayama2003}, the activation
energy is equal to the average kinetic energy of the condensing
particles minus the average energy they would have for $\beta =0$.
This gives
\begin{equation*}
E_{0}=\frac{\int_{0}^{\infty }d\text{v }\frac{1}{2}M\text{v}^{2}\text{ v }%
\left( 1-\beta \exp \left[ -\frac{M\text{v}^{2}}{2RT}\right] \right) \exp %
\left[ -\frac{M\text{v}^{2}}{2RT}\right] }{\int_{0}^{\infty }d\text{v v }%
\left( 1-\beta \exp \left[ -\frac{M\text{v}^{2}}{2RT}\right] \right) \exp %
\left[ -\frac{M\text{v}^{2}}{2RT}\right] }
\end{equation*}
\begin{equation}
-\frac{\int_{0}^{\infty }d\text{v }\frac{1}{2}M\text{v}^{2}\text{ v }\exp %
\left[ -\frac{M\text{v}^{2}}{2RT}\right] }{\int_{0}^{\infty }d\text{v v }%
\exp \left[ -\frac{M\text{v}^{2}}{2RT}\right] }=RT\frac{\beta }{2\left(
2-\beta \right) }  \label{5.7}
\end{equation}
Substitution of this equation in Eq.(\ref{5.6}) results in 
\begin{equation}
\sigma _{c}\left( T\right) =\left( 1-\frac{\beta }{2}\right) \exp \left( -%
\frac{\beta }{2\left( 2-\beta \right) }\right)   \label{5.8}
\end{equation}
which correponds to the result Nagayama and Tsuruta
\cite{Nagayama2003} found. These authors go on to give a further
expression for $\beta$ in terms of the system's specific volume
ratios.  One can also express the condensation coefficient in terms of
the activation energy:
\begin{equation}
\sigma _{c}\left( T\right) = \left( 1+\frac{2E_0}{RT}\right)^{-1}
\exp \left( {- \frac{E_0}{RT} }\right)   \label{5.9}
\end{equation} 


The flux does thus not follow a simple Arrhenius formula, not even for small
$E_0$. The deviation from Arrhenius behaviour is illustrated in
Fig.\ref{fig:3}. The figure shows that the value of the condensation
coefficient is always significantly smaller than expected from an
Arrhenious formula. Nagayama and Tsuruta found
$E_0/RT<1$.  The
importance of these expression is that they make it possible to
predict a priori the condensation coefficient (or the flux) from the
system's equilibrium properties. 

\section{Conclusions} 

In this paper we have proposed a new theory to deal with evaporation and
condensation phenomena in nonequilibrium situations. Our theoretical
framework is based on mesoscopic nonequilibrium thermodynamics \cite
{Vilar2001, Reguera1998}, a theory proposed to explore the mesoscopic level
of description, using the tenets of nonequilibrium thermodynamics.

We have derived an expression for the condensation flux, which can 
be applied to non-isothermal situations or even to the case in
which the system is away from mechanical equilibrium, when
acceleration of particles becomes significant. 

Our result reproduces the ones obtained using different approaches. In the
absence of a potential barrier along the condensation coordinate, we
rederive the expression for the rate given by the simplest version of
kinetic theory. Using the expression for the condensation probability
obtained by molecular dynamics of condensation, we obtain
the condensation coefficient found by Nagayama and Tsuruta
\cite{Nagayama2003}.  

Our theory has shown how the method, set up by nonequilibrium
thermodynamics \cite{deGroot1984} to deal with transport processes, also
holds for activated processes. In an early work of Eyring and Eyring 
\cite{Eyring1965} that possibility was questioned. An unified formalism,
able to cope with the difficulties inherent to nonequilibrium situations, has
then been established. 

In this sense, we can elucidate nonequilibrium situations in which the
system as a whole is subject to temperature and/or velocity gradients.
These are interesting generalisations of earlier conditions used. We
hope in this way to contribute to the understanding of phenomena which
pose interesting theoretical problems and are practically
important.

\section*{Acknowledgement} Discussions with T. Tsuruta are gratefully
acknowledged.

\newpage 

\section*{Figure Legends}

\vspace{1.5cm}

\mbox{}

 1. The interface separating the homogeneous gas and liquid phases on a $\mu$m-scale (a) and 
   in internal variable space (b).

\vspace{1.5cm}

2. The potential function
   $C^{s}\left(\gamma,\tilde{\mathrm{v}}\right)$ as a
   function of the 
   internal variables $\gamma$ and  $\tilde{\mathrm{v}}=\mathrm{v}/ (2RTM^{-1})^{-1/2}$.

\vspace{1.5cm}

3. The condensation coefficient as a function of the activation energy (solid line). The
   dotted line corresponds to Arrhenius behaviour.

\newpage

\vspace{15.5cm}

\newpage 

\vspace{1.5cm}

\newpage 

\begin{figure}
\begin{center}
\mbox{\psfig{file=evaporationfig3.eps,width=8cm}}
\caption{\label{fig:3}}
\end{center}
\end{figure}

\end{document}